\documentclass[showpacs,preprintnumbers,amsmath,amssymb]{revtex4}

\usepackage{graphicx} % Include figure files
\usepackage{amsmath}

\newenvironment{proof}[1][Proof]{\textbf{#1.} }{\ \rule{0.5em}{0.5em}}

\begin{document}

\title{MINIMAL GRAVITO-MAGNETISM}

\author{C.A. VAQUERA-ARAUJO. and J.L. LUCIO M.}
\email{vaquera@fisica.ugto.mx, lucio@fisica.ugto.mx}
\affiliation{\bigskip Instituto de F\'{i}sica de la Universidad de
Guanajuato, Loma del Bosque 103, Fracc. Lomas del Campestre, 37150
Le\'on, Guanajuato, M\'exico.}

\date{\today}

\begin{abstract}
\bigskip\noindent We show that Feynman's proof applies to Newtonian
gravitation, implying thus the existence of gravitational analogous
of the electric and magnetic fields and the corresponding
Lorentz-like force. Consistency of the formalism require particular
properties of the electric and magnetic-like fields under Galilei
transformations, which coincide with those obtained in previous
analysis of Galilean electromagnetism.
\end{abstract}


\maketitle

\section{\bigskip Introduction}

\noindent Historically the analogies between gravity and
electromagnetism have played an important role in the development of
Gravito-magnetism, even though is known that such analogies are
necessarily incomplete and therefore the natural framework where
this topic has developed is general relativity \cite{bahram}.
However it has been recognized that any theory including Newtonian
gravitation and Lorentz invariance in a consistent framework must
contain gravito-magnetism in some form \cite{bahramp} and even, a
close relation to Coriolis force has been remarked
\cite{Rindler,behera} . Thus, one wonders what is the minimal
framework where the phenomenon of gravitomagnetic forces occur. Our
interest in this short note is to point the relevance to
gravito-magnetism of what is known as Feynman`s proof \cite{Dyson}.
In fact, once one goes through the demonstration the conclusion is
evident, a particular form of gravito-magnetism follows just from
Galilei invariance and Newton`s second law.
 \bigskip

\noindent In 1990, F. Dyson \cite{Dyson} published an original proof
given by Feynman in 1948 of the homogeneous Maxwell equations
(divergenceless magnetic field and Faraday's law) and the Lorentz
Force Law. \ The motivation of Feynman was to discover a brand new
theory starting form simple assumptions, but the result was nothing
but the same old theory, and therefore, from his point of view, the
proof was more a failure than a success. Even though the proof is
mathematically correct, it requires some clarification
\cite{montesinos}. \ The proof is based on two essential parts: 1)
second Newton's law, 2) the commutator between components of the
position operator and between position and velocity. \ There is an
apparent inconsistency in these assumptions since the first is
purely classical while the second comes from a quantum theory,
however Bracken\cite{Bracken} remarked that it is possible to
substitute the quantum commutators by classical Poisson brackets.
\bigskip

\noindent According to our view, the logic of the proof is the
following.

\begin{itemize}
    \item Galilean relativity is the basis of the formulation.
    \item Galilean invariance is enough to derive minimal
    coupling, {\it i.e.} to introduce electromagnetic
    interactions.
    \item Newton's second law and minimal coupling are consistent
    with the Lorentz force and two homogeneous Maxwell equations,
    provided the electric and magnetic field transforms
    appropriately under boosts.
\end{itemize}

\noindent It is important to remark  that, in the second and third
points above, we can change the electromagnetic interactions --
Maxwell equations and Lorentz force -- by more generic terms since
no where in the proof the electromagnetic nature of the vector
potential is invoked. That has motivated us to pursue the
implications of Feynman's proof to the gravitomagnetic interaction,
a point that seems to be overlooked so far \cite{ref}.

\bigskip

\noindent In order to introduce the notation and for easy of
reading, in this short note we first formulate Feynman's proof and
summarize the ingredients of the Galilei group required for the
presentation. With these tools at hand we show that the assumptions
required in Feynman's proof can be derived from the Galilei algebra;
in particular, minimal coupling is discussed following the approach
by Levy-Leblond \cite{Levy}. Requiring consistency of the whole
approach we derive the properties of the electric and magnetic
fields under boost transformations, a point that deserves attention
since the non-relativistic limit of the transformation of the fields
is ambiguous, a point discussed long time ago \cite{LeBel}. Finally
the applicability of Feynman`s proof to gravito-magnetism is
discussed, in particular the conditions for its validity.
\bigskip

\subsection{Feynman's proof}

\bigskip

\noindent $\textsl{Assume}$ a particle exists with position
$x_{j}$ ($j$=1,2,3) and velocity $\dot{x}_{j}$ satisfying Newton's
Second Law
\begin{equation}
m\ddot{x}_{j}=F_{j}(x,\dot{x},t), \label{f1}%
\end{equation}
\noindent with Poisson brackets
\begin{eqnarray} \label{f2}
\{x_{i},x_{j}\}  &  =0
\end{eqnarray}
\begin{eqnarray} \label{f3}
m\{x_{i},\dot{x}_{j}\}  &  =\delta_{ij}.
\end{eqnarray}

\noindent \textsl{Then} there exists \ fields $\mathbf{E}(x,t)$ and $\mathbf{B}%
(x,t)$ satisfying the Lorentz force and Maxwell equations
\begin{equation}
F_{j}=E_{j}+\epsilon_{jkl}\dot{x}_{k}B_{l}, \label{f4}%
\end{equation}

\begin{equation}
\mathbf{\nabla\cdot B}=0, \label{M1}%
\end{equation}%
\begin{equation}
\frac{\partial\mathbf{B}}{\partial t}+\mathbf{\nabla\times E}=0. \label{M2}%
\end{equation}

\begin{proof}
\noindent {From} Eqs. (\ref{f1},\ref{f3}) it follows:
\begin{equation}
\{x_{j},F_{k}\}+m\{\dot{x}_{j},\dot{x}_{k}\}=0. \label{f5}%
\end{equation}

\noindent The Jacobi identity%
\begin{equation} \label{e8}
\{x_{l},\{\dot{x}_{j},\dot{x}_{k}\}\}+\{\dot{x}_{j},\{\dot{x}_{k}%
,x_{l}\}\}+\{\dot{x}_{k},\{x_{l},\dot{x}_{j}\}\}=0,
\end{equation}
together with Eq. (\ref{f3}) and Eq. (\ref{f5}) imply
\begin{equation}
\{x_{l},\{x_{j},F_{k}\}\}=0, \label{f7}%
\end{equation}
while Eq. (\ref{f5}) allow us to conclude that
\begin{equation}
\{x_{j},F_{k}\}=-\{x_{k},F_{j}\},
\end{equation}
and therefore, we can write
\begin{equation}
\{x_{j},F_{k}\}=-\frac{1}{m}\epsilon_{jkl}B_{l}. \label{b1}%
\end{equation}
Eq. (\ref{b1}) is the definition of $\mathbf{B}$, which by virtue
of Eq. (\ref{e8}) can be written as
\begin{equation}
B_{l}=\frac{m^{2}}{2}\epsilon_{jkl}\{\dot{x}_{j},\dot{x}_{k}\} \label{b2}%
\end{equation}
On the other hand Eq. (\ref{f7}) can also be expressed as follows
\begin{equation}
\{x_{j},B_{l}\}=0, \label{f8}%
\end{equation}
which means that $\mathbf{B}$ is only a function of $\mathbf{x}$
and $t$.

\bigskip
\noindent Defining $\mathbf{E}$ by Eq. (\ref{f4}), which guarantee
the Lorentz Force is correctly incorporated, and using Eqs.
(\ref{f3}, \ref{b1}) and Eq. (\ref{f8}), it follows that
$\mathbf{E}$ is  only a function of $\mathbf{x}$ and $t$.
\begin{equation}
\{x_{j},E_{l}\}=0, \label{f9}%
\end{equation}
Moreover, using the expression for $\mathbf{B}$, Eq. (\ref{b2})
the Jacobi identity :
\begin{equation}
\epsilon_{jkl}\left\{  \dot{x}_{l},\{\dot{x}_{j},\dot{x}_{k}\}\right\}  =0.
\label{f10}%
\end{equation}
can be cast in the form:
\begin{equation}
\{\dot{x}_{l},B_{l}\}=0, \label{f11}%
\end{equation}
which is equivalent to the Maxwell equation (\ref{M1}).

\bigskip

\noindent The time evolution of $\mathbf{B}$ is obtained from the
time derivative of Eq. (\ref{b2}). \ This gives :
\begin{equation}
\frac{\partial B_{l}}{\partial t}+\frac{\partial B_{l}}{\partial x_{m}%
}\dot{x}_m =m^2\epsilon_{jkl}\{\ddot{x}_{j},\dot{x}_{k}\}. \label{f12}%
\end{equation}
Now by Eq. (\ref{f1}) and Eq. (\ref{f4}), Eq. (\ref{f12}) becomes%
\begin{align}\label{f18}
\frac{\partial B_{l}}{\partial t}&+\dot{x}_{m}\frac{\partial B_{l}}%
{\partial x_{m}}=m\epsilon_{jkl}\{E_{j}+\epsilon_{jmn}%
\dot{x}_{m}B_{n},\dot{x}_{k}\}\\
& =  m\left(  \epsilon_{jkl}\{E_{j},\dot{x}_{k}\}+\{\dot{x}_{k}B_{l},\dot{x}%
_{k}\}-\{\dot{x}_{l}B_{k},\dot{x}_{k}\}\right) \nonumber\\
& =\epsilon_{jkl}\frac{\partial E_{j}}{\partial x_{k}}+\dot{x}_{k}%
\frac{\partial B_{l}}{\partial x_{k}}-\dot{x}_{l}\frac{\partial B_{k}%
}{\partial x_k} -mB_{k}\{\dot{x}_{l},\dot{x}_{k}\}.\nonumber
\end{align}\
\noindent Using Eq. (\ref{b2}) one shows the last term is zero by
symmetry while the third term vanishes because of Eq. (\ref{f11}).
Thus:
\begin{equation}
\frac{\partial B_{l}}{\partial t}=\epsilon_{jkl}\frac{\partial E_{j}}{\partial
x_{k}},
\end{equation}
which is equivalent to Eq(\ref{M2}).
\bigskip
End of proof.
\end{proof}

\bigskip

\section{Galilei group in 3+1 D}

\bigskip

\noindent The three dimensional Galilei group is defined as the
ten parameter Lie group of the space time transformations of the
form:
\begin{equation}
\genfrac{}{}{0pt}{}{\mathbf{x}^{\prime}=\mathcal{R(}\theta,\varphi
,\psi\mathcal{)}\mathbf{x}+\mathbf{v}t+\mathbf{u}}{t^{\prime}=t+\tau}%
.\label{transf1}%
\end{equation}

\noindent where $\mathcal{R}$ is a SO(3) rotation matrix. The Lie
algebra of the three dimensional Galilei group is ordinarily
referred to a conventional basis consisting of ten generators: time
and space translations $H, P_i$, rotations $J_i$, and boosts $K_i. $
It is well known that Galilei group possess a family of nontrivial
projective representations \cite{Levy} characterized by a real
number $m$, which in physical systems is interpreted as the particle
mass. The corresponding Poisson brackets are:
\bigskip

\begin{equation} \label{gal}%
\begin{array}{cc}
\{  J_{i},J_{j}\} = \epsilon_{ijk}J_{k},~~ \{J_{i},K_{j}\}  =
\epsilon_{ijk}K_{k},~~ \{  H,P_{j}\}  =0,~~ \{J_{i},P_{j}\}
=\epsilon_{ijk}P_{k},\\ \{ H,J_{i}\} = 0, ~~ \{H,K_{j} \}=-P_{j},
~~\{K_{i},K_{j}\} = 0,~~\{P_{i},P_{j}\}=0,~~  \{ K_{i},P_{j}\}
=m\delta_{ij}.
\end{array}
\end{equation}
\bigskip

\noindent The localization properties of the system can be
investigated looking for a position function $x_{i}$ $(i=1,2,3)$ in
the enveloping Galilei Lie algebra. \ The natural requirements
$x_{i}$\ must obey, in order to be identified with the spatial
position, are \cite{Levy}:

\begin{enumerate}
\item  A state localized at $x_{i}$ transforms under a translation
by $u_{i}$ in a state localized at $x_{i}+u_{i}$. \ In other
words, it requires the validity of the Poisson bracket rule
\begin{equation}
\{  x_{i},P_{j}\}  =\delta_{ij}, \label{pos1}%
\end{equation}

\item  it should transform like a vector under spatial rotations, or
equivalently,
\begin{equation}
\{  J_{i},x_{j}\}  =\epsilon_{ijk}x_{k}. \label{pos2}%
\end{equation}

\item  An instantaneous ($t=0$) boost, must leave invariant the
position, \it{i.e.}
\begin{equation}
\left.  \{  K_{i},x_{j}\}  \right|  _{t=0}=0. \label{pos3}%
\end{equation}
\end{enumerate}

\bigskip These conditions are fulfilled by the following function:
\begin{equation}
x_{i}=\left.  \frac{K_{i}}{m}\right|  _{t=0}. \label{pos4}%
\end{equation}
Once the relation between the boost generator and the position is
established, it is clear that the mass as a central extension
--- last relation in Eq. (\ref{gal})--- plays an important role
in the classical relation that is used in \cite{Bracken} to replace
the quantum commutator in Feynman's assumptions \cite{Dyson}.
\bigskip

\section{Feynman's proof and galilean invariance}

\bigskip

\noindent We are now ready to analyze the hypotheses of Feynman's
proof with to the light of galilean relativity. \ The second
assumption Eq. (\ref{f2}) is an immediate consequence of Eqs.
(\ref{pos3},\ref{pos4}), which define the action of instantaneous
boosts (at $t=0$) on the
particle position:%
\begin{equation}
\left.  \left\{  x_{i},\frac{k_{j}}{m}\right\}  \right|  _{t=0}=\left\{
x_{i},x_{j}\right\}  =0.
\end{equation}

\bigskip

\noindent The first assumption, Eq. (\ref{f1}), involves in fact
two relations:

\begin{itemize}
\item  Newton's second law %
\begin{equation}
F_{i}=\frac{d\pi_{i}}{dt}\qquad(i=1,2,3),
\end{equation}
where $\pi_{i}\ (i=1,2,3)$ is the kinematical momentum of the
particle, not necessarily equal to the canonical momentum $p_{i}$,

\item  and the statement that $\pi_{i}$ is related to the velocity
of the particle $\dot{x}_{i}$ according
to%
\begin{equation}
F_{i}=\frac{d\pi_{i}}{dt}=m\frac{d\dot{x}_{i}}{dt}\qquad(i=1,2,3),
\end{equation}
or equivalently%
\begin{equation}
\pi_{i}=m\dot{x}_{i}+c_{i},
\end{equation}
where $c_{i}$ is a constant vector that can be absorbed into the
definition of $\pi_{i}$:
\begin{equation}
\pi_{i}=m\dot{x}_{i}. \label{km}%
\end{equation}
\end{itemize}

\bigskip

\noindent Notice that the relation between velocity and
kinematical momentum
Eq. (\ref{km}) severely restricts the form of the Hamiltonian since we have%
\begin{equation}
\pi_{i}=m\left\{  x_{i},H\right\}.
\end{equation}

\bigskip

\noindent The connection of the third assumption Eq. (\ref{f3})
with Galilei algebra goes through the relation among velocity and
momentum; therefore in order to proceed, we need to know in first
place the relation between canonical $p_{i}$ and kinematical
momentum $\pi_{i}$.\ The following argument due to L\'{e}vy-
Leblond \cite{Levy}, provides the desired link. \ Indeed, it is
enough to demand the existence of instantaneous boost
transformations of
momentum and position%
\begin{align}
p_{i}  &  \rightarrow p_{i}+mv_{i}\label{int1}\\
x_{i}  &  \rightarrow x_{i},\nonumber
\end{align}
and to postulate that the kinematical momentum transforms in the
same way, not only for the free particle, but also when
interactions are introduced
\begin{equation}
\pi_{i}\rightarrow\pi_{i}+mv_{i}. \label{int2}%
\end{equation}
In this case, the transformation of $\pi_{i}$ is noting but the
familiar velocity composition under a boost. As required, Eqs.
(\ref{pos1},\ref{pos2},\ref{pos3}) remain valid, and therefore,
comparing Eq. (\ref{int1}) and Eq. (\ref{int2}), we conclude that
under a boost:
\begin{equation}
p_{i}-\pi_{i}\rightarrow p_{i}-\pi_{i},
\end{equation}
thus, the functions $A_{i}=p_{i}-\pi_{i}$ \ $(i=1,2,3)$\ satisfy%
\begin{equation}
\left.  \left\{  k_{i},A_{j}\right\}  \right|  _{t=0}=m\left\{  x_{i}%
,A_{j}\right\}  =m\frac{\partial A_{j}}{\partial p_{i}}=0,
\end{equation}
that is, $\mathbf{A}$ is a function of $\mathbf{x}$ alone (and possibly of
time). \ Then, the relation between the canonical momentum and the kinematical
momentum is, using Eq. (\ref{km}),%
\begin{equation}
p_{i}=\pi_{i}+A_{i}(x,t)=m\dot{x}_{i}+A_{i}(x,t). \label{mc}%
\end{equation} \

\noindent This is nothing but minimal coupling, which has been
obtained from Galilei relativity plus plausible assumptions (a
formal proof of the derivation of minimal coupling based on
Galilei relativity can be found in \cite{jauch}). Now the third
assumption, Eq. (\ref{f3}), may be seen as a
consequence of Eqs. (\ref{pos1},\ref{mc}).%
\begin{align}
\left\{  x_{i},p_{j}\right\}  &=\left\{  x_{i},m\dot{x}_{j}\right\}  +\left\{
x_{i},A_{j}(\mathbf{x},t)\right\}\\
&=m\left\{  x_{i},\dot{x}_{j}\right\}
=\delta_{ij}\nonumber.
\end{align}

\bigskip

\noindent So far we have shown that besides Newton's second Law,
the hypothesis used by Feynman follow from the 3+1 dimensional
Galilei algebra. In order to check the consistency of the output
we now investigate the transformation laws for the Electric and
Magnetic fields. Since the properties of the differential
operators under boosts follow from Eqs. (\ref{transf1},\ref{int1})

\begin{equation}
\begin{array}
[c]{c}%
\frac{\partial}{\partial x_{i}^{\prime}}=\frac{\partial}{\partial x_{i}}\\
\frac{\partial}{\partial t^{\prime}}=\frac{\partial}{\partial t}-v_{i}%
\frac{\partial}{\partial x_{i}}\\
\frac{\partial}{\partial p_{i}^{\prime}}=\frac{\partial}{\partial p_{i}}.
\end{array}\label{dop}
\end{equation} \
\noindent then, the Lorentz force Eq. (\ref{f4}) transforms as:
\begin{equation}
F_{i}^{\prime}=\frac{d\pi_{i}^{\prime}}{dt^{\prime}}=\frac{d\pi_{i}}{dt}%
=F_{i}.
\end{equation}

\bigskip

\noindent On the other hand, according to the definition of
$\mathbf{B}$ and \textbf{$\mathbf{\pi}$}, Eqs.
(\ref{b2},\ref{km}):
\begin{equation}
B_{l}=\frac{1}{2}\epsilon_{jkl}\{\pi_{j},\pi_{k}\},
\end{equation}
then, the transformation law for the magnetic field is%
\begin{align}
B_{l}^{\prime} &=\frac{1}{2}\epsilon_{jkl}\{\pi_{j}^{\prime},\pi_{k}^{\prime
}\}^{\prime}\\
%&=\frac{1}{2}\epsilon_{jkl}\left(\frac{\partial \pi_{j}^{\prime}}{\partial x_{i}^{\prime}}
%\frac{\partial \pi_{k}^{\prime}}{\partial p_{i}^{\prime}}-
%\frac{\partial \pi_{j}^{\prime}}{\partial p_{i}^{\prime}}
%\frac{\partial \pi_{k}^{\prime}}{\partial x_{i}^{\prime}}\right)\nonumber\\
%&=\frac{1}{2}\epsilon_{jkl}\left(\frac{\partial \pi_{j}}{\partial x_{i}}
%\frac{\partial \pi_{k}}{\partial p_{i}}-
%\frac{\partial \pi_{j}}{\partial p_{i}}
%\frac{\partial \pi_{k}}{\partial x_{i}}\right)\nonumber\\
&=\frac{1}{2}\epsilon_{jkl}\{\pi_{j},\pi_{k}\}=B_{l}.\nonumber
\end{align}

\bigskip

\noindent Finally, the transformation of the Electric field, under boosts is%
\begin{align}
E_{i}^{\prime}&=F_{i}^{\prime}-\epsilon_{ikl}\frac{\pi_{k}^{\prime}}{m}%
B_{l}^{\prime}=F_{i}-\epsilon_{ikl}\frac{\left(  \pi_{k}+mv_{k}\right)  }%
{m}B_{l}\\
&=E_{i}-\epsilon_{ikl}v_{k}B_{l}.\nonumber
\end{align}
Thus, Lorentz force and the two homogeneous Maxwell equations, are
consistent with Galilean relativity if the electric and magnetic
field transform according to:
\begin{equation} \label{t3}
\begin{array}
[c]{c}%
\mathbf{E}^{\prime}=\mathbf{E-v\times B,}\\
\mathbf{B}^{\prime}=\mathbf{B}.
\end{array}
\end{equation}

\bigskip

\noindent Can these transformation properties be identified with
the non-relativistic limit of Maxwell equations ? It turns out
\cite{LeBel} that two such limits exist (these can be traced back
to the relation $\epsilon_0\mu_0c^2=1$, since in the $c \to
\infty$ limit $\epsilon_0$ and $\mu_0$ can not remain finite
simultaneously). In none of these non-relativistic limits Eqs.
(\ref{M1}, \ref{M2}) \textbf{and} the Lorentz force Eq. (\ref{f4})
can be obtained together with the transformation rules Eq.
(\ref{dop}) \textbf{and} Eq. (\ref{t3}).  Thus, the transformation
properties of the fields Eq. (\ref{t3}) can not be obtained from a
non-relativistic limit; therefore care must be exercised when
considering the non-relativistic theory of Maxwell equations,
which should be defined not only as the $c$ large limit, but also
as the limit that ensures the correct transformation properties
under the Galilei group.

\section{\bigskip gravito-magnetism}

\noindent The relation to gravito-magnetism arises from the
observation that the vector potential involved in the derivation
of minimal coupling, as presented in the previous section or in
Ref. \cite{jauch}, has nothing to do with electromagnetism, is
valid in more general grounds. In fact the same proof could lead
to Newtonian gravitation, therefore gravito-magnetism should also
be derived solely from Newton's equation of motion and Galilei
invariance \cite{ref}.
\bigskip

\noindent The derivation of Maxwell-type gravitational equations
and Lorentz-like force has a long history \cite{weinberg, mashh}.
These relations, and others applying in the relativistic domain,
are usually derived starting with the gravitational field
equations.  Here we restrict our attention to Electromagnetic-like
effects of a stationary space-time in the low velocity and weak
field approximation.
\bigskip

\noindent The characteristic features of a stationary space-time
are the following \cite{Rindler}:
\begin{itemize}
\item its metric tensor is independent of time, that is
\begin{equation}
ds^{2}=g_{\mu\nu}(x_{i})dx^{\mu}dx^{\nu},(\mu,\nu=0,\cdots,3;
i=1,\cdots,3).
\end{equation}

\item it is always possible to find a canonical form of the metric
in which, according to time dilation, the component $g_{00}$ can
be parameterized as %
\begin{equation}
g_{00}=e^{\frac{2\Phi(x)}{c^{2}}},%
\end{equation}
where $\Phi(x)$ is the clock-rate function (in the approximation we work,
the gravitational potential).%
\end{itemize}

\bigskip

\noindent The so-called canonical form of the stationary
space-time metric is, then,
\begin{equation}\label{met1}
ds^{2}=e^{2\Phi(x)/c^{2}}\left(
cdt-\frac{1}{c^{2}}w_{i}(x)dx^{i}\right)
^{2}-k_{ij}(x)dx^{i}dx^{j},
\end{equation}
where $w_{i}(x)$ and $k_{ij}(x)$ are time independent
coefficients.  It can be shown that under the transformation of
the time coordinate of the stationary metric:
\begin{equation}
t\rightarrow \kappa \left[t+f(x)\right],
\end{equation}
Equation (\ref{met1}) remain invariant provided the functions
$\Phi$, $w_i$ and $k_{ij}$ transform as follows:
\begin{equation}
\begin{array}
[c]{c}%
\Phi \rightarrow \Phi-{c^{2}}\ln\kappa \\
w_{i} \rightarrow \kappa \left(w_{i}+{c^{3}}\frac{\partial
f}{\partial x^{i}}\right)\\
k_{ij} \rightarrow k_{ij}.
\end{array}
\end{equation}

\bigskip

\noindent In the slow-motion and weak field approximation, the
metric Eq. (\ref{met1}) can be replaced by
\begin{align}
ds^{2}  & =\left(  1+\frac{2\Phi(x)}{c^{2}}\right)  \left(  1-\frac{1}{c^{3}%
}w_{i}\frac{dx^{i}}{dt}\right)  c^{2}dt^{2}-d\mathbf{x}^{2}\\
& \simeq\left(
1+\frac{2\Phi(x)}{c^{2}}-\frac{1}{c^{3}}\mathbf{w\cdot
\dot{x}}\right) c^{2}dt^{2}-d\mathbf{x}^{2}.\nonumber
\end{align}
Then, the action of a massive particle that moves between points
$P_{1}$ and $P_{2}$ ---under de action of a weak gravitational
field--- is
\begin{equation}
S\left[t,\mathbf{x}(t)\right]=-m\int_{P_{1}}^{P_{2}}ds\simeq-m_{0}c\left\{  t_{2}-t_{1}-\frac{1}{c}%
\int_{P_{1}}^{P_{2}}\left[  \frac{\dot{x}^{2}}{2}-\Phi(x)-\frac{\mathbf{w\cdot \dot{x}}%
}{c}\right]  dt\right\}  .
\end{equation}
The corresponding variational principle yields the Lorentz-type
force equation
\begin{equation}\label{lf1}
\mathbf{f}=m\mathbf{\ddot{x}}=-m\mathbf{\nabla}\Phi+\frac{m}{c}\mathbf{\dot{x}}\times\left(
\mathbf{\nabla}\times\mathbf{w}\right).
\end{equation}

\bigskip

\noindent $\mathbf{w}$ is subject to the condition
$\partial\mathbf{w}/\partial t=0$ and plays the role of a
gravito-magnetic vector potential, which turns out to be related
to the local rotation rate of the reference frame. \ From the
point of view of Newtonian mechanics, if a point $P$ of a rigid
reference frame $L$ travels with acceleration
$\mathbf{a}={\mathbf{E}}/{m}$ through an inertial frame while $L$
rotates about $P$ at angular velocity
$\mathbf{\Omega}={\mathbf{B}}/{2 m}$, then a free particle of mass
$m$ at $P$ moving relative to $L$ at velocity $\mathbf{\dot{x}}$
experiences a force
\begin{equation}\label{lf2}
\mathbf{f}=m\mathbf{\ddot{x}}=\mathbf{E}+\mathbf{\dot{x}}\times{\mathbf{B}},
\end{equation}
where the last term corresponds to the well known Coriolis force.
Comparison of Eqs. (\ref{lf1}) and (\ref{lf2}) makes it possible
to establish  the relations:
\begin{equation}
\mathbf{E(x)}=-m\mathbf{\nabla}\Phi(x)
\end{equation}
\begin{equation}
\mathbf{B(x)=}\frac{m}{c}\left[
\mathbf{\nabla}\times\mathbf{w}(x)\right],
\end{equation}
therefore, by construction, the fields $\mathbf{E}(x)$ and
$\mathbf{B}(x)$ satisfy
\begin{equation}
\mathbf{\nabla\cdot B}=0, \label{M21}%
\end{equation}%
\begin{equation}
\mathbf{\nabla\times E}=0. \label{M22}%
\end{equation}

\noindent Thus, if we suppose that the force does not depend
explicitly on time, {\it i.e.} if we replace the assumption Eq.
(\ref{f1}) by the more restrictive one
\begin{equation}
m\ddot{x}_{j}=F_{j}(x,\dot{x}), %
\end{equation}
\noindent then Feynman's proof is able to reproduce the
stationary, weak field and low velocity partial description of
gravito-magnetism, given by Eqs(\ref{lf2}, \ref{M21} and
\ref{M22}). In this case again, the argument has nothing to say
regarding the relation between the fields $\mathbf{B}(x)$ and
$\mathbf{E}(x)$ and the sources (the complementary inhomogeneous
Maxwell-like gravito-magnetic equations).

\bigskip

\noindent Finally it is important to remark that in the
\textbf{non-stationary} case Feynman's proof is incompatible with
gravito-magnetism, because in such a case \cite{mashh}, Eq.
(\ref{M22}) is not replaced by Eq. (\ref{M2}) but instead by
\begin{equation}
\frac{1}{4}\frac{\partial\mathbf{B}}{\partial t}+\mathbf{\nabla\times E}=0. \label{M3}%
\end{equation}

\section{\bigskip Discussion}

In this paper we have shown that: \bigskip

\begin{itemize}
    \item The assumptions used in Feynman's proof are either
    consistent with Galilei invariance (Newton's second law),
    or derived from it (Eqs. (\ref{f2},\ref{f3})).
    \item Minimal coupling is derived from Galilei invariance.
    Although a formal proof of this is given in \cite{jauch}, here
    we presented the argument following Levy-Leblond \cite{Levy}.
    \item Appropriated transformations under boosts of the
    electric and magnetic fields exist such that the Lorentz force
    and Maxwell equations are consistent with Galilean relativity.
    \item  Feynman's proof is able to reproduce the stationary, weak
    field and slow-motion approximation of gravito-magnetism,
    assuming the fields do not depend explicitly on time.
\end{itemize}

\noindent A question that comes to mind immediately is if the
derivation of minimal coupling (gauge principle) starting from the
Galilei group can be extended to the relativistic domain. As far as
we know all attempts in this direction have failed \cite{jauch}, for
they cannot incorporate the inherent reparametrization invariance of
the relativistic theory \cite{Tanimura}.
\bigskip

\noindent We conclude that Feynman's proof is valid in the framework
of Galilean relativity (Dyson's statement referring to
Feynman's proof \cite{Dyson}
     "The proof begins with assumptions
invariant under Galilean transformations and ends with equations
invariant under Lorentz transformations"  turns out to be
incorrect) and that  Feynman's proof applies to Newtonian
gravitation, implying thus the existence of gravitational
analogous of the electric and magnetic fields and the
corresponding Lorentz-like force.

\begin{acknowledgments}
\noindent The authors acknowledge financial support from SNI,
CONACyT--M\'exico under project 44644-F and Abdus Salam ICTP
Associateship scheme. \ J.L. Lucio M. also acknowledges support from
CONCyTEG (1417-027) and DINPO--UGTO.
\end{acknowledgments}

\end{document}